\documentclass[aps,a4paper,showpacs,twocolumn]{revtex4}
\usepackage{graphicx}
\usepackage{amsmath}
\usepackage{amssymb}
\usepackage{enumerate, color}
\usepackage{subfigure}
\usepackage{tabularx}
\usepackage{epstopdf}
\usepackage{lipsum}
\newcommand{\be}{\begin{equation}}
\newcommand{\ee}{\end{equation}}
\newcommand{\ben}{\begin{eqnarray}}
\newcommand{\een}{\end{eqnarray}}
\newcommand{\bes}{\begin{subequations}}
\newcommand{\ees}{\end{subequations}}

\newcommand{\bb}{\bibitem}

\begin{document}
\title{From sine-Gordon to vacuumless systems in flat and curved spacetimes}
\author{D. Bazeia}
\author{D.C. Moreira}
\affiliation{Departamento de F\'isica, Universidade Federal da Para\'iba, 58051-970, Jo\~ao Pessoa, PB, Brazil}
\begin{abstract}
In this work we start from the Higgs prototype model to introduce a new model, which makes a smooth transition between systems with well located minima and systems that support no minima at all. We implement this possibility using the deformation procedure, which allows the obtention of a sine-Gordon-like model, controlled by a real parameter that gives rise to a family of models, reproducing the sine-Gordon and the so-called vacuumless models. We also study the thick brane scenarios associated with these models and investigate their stability and renormalization group flow. In particular, one shows how gravity can change from the 5-dimensional warped geometry with a single extra dimension of infinite extent to the conventional 5-dimensional Minkowski geometry. 
\end{abstract}
\maketitle
\section{Introduction} 

Topological solutions in Field Theory are related to many phenomena in Physics \cite{vachaspati,v,ms}. In particular, when we have models in 1+1 dimensions involving a potential written in terms of a scalar field and with a set of degenerate minima, such solutions represent transitions between consecutive minima and are called {\it kinks}. The minima define the possible vacua states at the quantum level, and are distributed over the many values ​​for the field. Such distribution can appear in the most diverse ways, for instance, in the $\phi^4$-model one has only one topological sector, defined between two consecutive minima, while in the sine-Gordon model \cite{caudrey} one has an infinite copy of the same sector, always between two consecutive and well localized minima. A system that is quite different from this perspective is the vacuumless model \cite{vilenkin1,vilenkin2}. In this model we still have a topological sector, which can be interpreted as connecting two minima of the scalar potential, but now they are located at infinity. In this case, the field solution is asymptotically divergent and has infinite amplitude, but keeps its topological character well behaved \cite{bvacless}. It is worth mentioning that vacuumless systems appear in a diversity of contexts in high energy physics \cite{vilenkin1,vilenkin2,bvacless,bbn,bbl,bbg,dutra}.

The aim of this paper is to construct a model that, with the proper variation of a given parameter, makes the transition between systems with well-located minima and systems that support no minima at all. The model presented here has as limit cases the sine-Gordon and the vacuumless models. Using the deformation procedure developed in Ref.~\cite{blm}, we find a field transformation that takes us to a new system with a double sine-Gordon-like behavior, which has two infinitely degenerate sets of solutions. One of these sets transits between the sine-Gordon kink and the vacuumless solution. The second set coincides with the previous one in one of its limits and, except from a phase, in the other limit case it is destroyed, remaining only the zero energy solutions. Thus, in one of the limits of the model the two phases coincide, generating infinite copies of the same topological sector, and in the other limit only one topological sector survives.

Models in Field Theory motivate generalizations of the Randall-Sundrum model \cite{rs1,rs2} in the presence of scalar fields \cite{gw,fre,csaki,gremm,d,bm,bmm}, which are known as {\it thick branes}. For this reason, in this work we also study the thick brane scenario generated by the scalar field model which we first introduce and study in the flat spacetime. In particular, we find a brane that transits between the sine-Gordon brane \cite{gremm} and a flat 5-dimensional spacetime with two zero energy solutions for the scalar field. The idea here is similar to the case investigated in \cite{bc}, in which the authors propose a braneworld scenario where the brane changes from a thick to a thin behavior. In the current study, however, we describe a mechanism in which a single parameter can be used to control the brane profile, contributing to change the 5-dimensional warped geometry into a flat geometry.

Although the braneworld model that we explore below is more involved, it also supports analytical solutions. Thus, in the 5-dimensional spacetime with a single extra dimension of infinite extent, we also analyze the stability of the braneworld scenario against tensorial perturbations. In the sense of Gauge/Gravity Duality \cite{maldacena, skenderis1}, where the extra dimension can be identified with the energy scale of the holographic dual field theory, we also study its implications for the renormalization group flow (the RG flow) in the dual Field Theory, since stability of the gravitational sector has relevant information about the dual model \cite{skenderis}.

The subject to be explored in the current work is organized as follows. In Sec. II  we review several aspects of the first order formalism for a single real scalar field and its relationship with the Bogomol'nyi-Prasad-Somerfield (BPS) solutions \cite{bps}. Then, we discuss the deformation procedure and show how it acts to generate the new model in the 2-dimensional spacetime. We go on and study in Sec. III the new model and its topological solutions, including how the energy and energy density behave. In addition, we also show that the solutions obtained are stable. Moreover, in Sec. IV we analyze the properties of the thick branes that can be constructed from the model. We also investigate stability of the brane against metric fluctuations and study implications of the RG flow for the dual Field Theory. We end the work in Sec.~\ref{end}, adding some comments and conclusions.

\section{Generalities}

\subsection{First order formalism}

The  behavior of a scalar field is usually encoded in a Lagrangian density having the general form
\be\label{ori}
\mathcal{L}(\phi,\partial_{\mu}\phi)=\frac{1}{2}\partial_{\mu}\phi\partial^{\mu}\phi-V(\phi).
\ee
Here $\phi$ is the scalar field and $V(\phi)$ is the potential of the model, which  determines how the field behaves. In the flat spacetime with $(1,1)$ spacetime dimensions, the metric tensor becomes $\eta_{\mu\nu}=\text{diag}(1,-1)$, so the scalar field only depends on the two coordinates $x$ and $t$, i.e., $\phi=\phi(x,t)$. For simplicity, we also work with dimensionless fields and coordinates. The equation of motion for the scalar field derived from the Lagrangian (\ref{ori}) is given by 
\begin{equation}\label{equationofmotion}
\partial_{\mu}\partial^{\mu}\phi+\frac{dV}{d\phi}=0.
\end{equation}
As the Lagrangean (\ref{ori}) is Lorentz invariant, we can focus on static solutions, since traveling waves can be obtained from a Lorentz boost.  For static configurations the equation (\ref{equationofmotion}) becomes a second order differential equation given by
\begin{equation}\label{2ode}
\frac{d^2\phi}{dx^2}=\frac{dV}{d\phi}.
\end{equation}
Another important quantity we are interested in is the energy-momentum tensor 
\be\label{emt}
T^{\mu\nu}=\partial^{\mu}\phi\partial^{\nu}\phi-\eta^{\mu\nu}\mathcal{L}.
\ee 
In particular, its $T^{00}$ component provides the energy density of the solution we are looking for. We represent it by $\rho(x)$, which is explicitly given by
\bes\label{rho1}\ben
\rho (x)&=&\frac{1}{2}\phi'^2+V(\phi)\\
&=&\frac{1}{2}\left(\phi'\mp\sqrt{2V(\phi)}\right)^2\pm\phi'\sqrt{2V(\phi)}.
\een\ees
Note that for positive-definite energy, the potential must be non-negative, i.e., $V(\phi)\geq 0$. A powerful tool in the treatment of these models is the use of an auxiliary function, denoted by $W(\phi)$, which is introduced as follows
\begin{equation}\label{potv}
V(\phi)=\frac{1}{2}W_\phi^2,
\end{equation}
where $W_{\phi}={dW}/{d\phi}$. In this case the expression for the energy density (\ref{rho1}) becomes
\begin{equation}\label{rho2}
\rho (x)=\frac{1}{2}\left(\phi'\mp W_{\phi}\right)^2\pm\frac{dW}{dx}
\end{equation}
and the equation (\ref{2ode}) can now be given as the first order differential equations
\begin{equation}\label{1ode}
\frac{d\phi}{dx} =\pm W_{\phi}.
\end{equation}
It implies that the quadratic term in (\ref{rho2})  disappears and, as a consequence, the energy density is only related to the $x$-derivative of  $W$. Thus, the energy of the model is determined only by the asymptotic behavior of the function $W$ in the coordinate space; that is, one can write
\begin{eqnarray}\label{Ebps}
E_{BPS}&=&\int_{-\infty}^{\infty}\rho(x)\, dx\nonumber\\
&=&|W\left(\phi (\infty)\right)\!-\!W\left(\phi \left(-\infty)\right)\right)|.
\end{eqnarray}
In this case the energy (\ref{Ebps}) is called BPS energy \cite{bps}.

Relevant phenomena occurs when the system under analysis presents a set of degenerate minima. In these situations each pair of consecutive minima form distinct topological sectors that, in turn, have different solutions. These solutions are called {\it kinks}. The simplest case of models having such properties is the well-known Higgs prototype or $\phi^4$-model, defined by the potential
\begin{equation}\label{phi4}
V(\phi)=\frac{1}{2}(1-\phi^2)^2,
\end{equation}
which has two degenerate minima at $\phi=\pm 1$ and a topological sector having  a kink solution given explicitly by $\phi(x)=\tanh(x)$.

An interesting way to characterize a kink is the existence of a topological current. Here, we define it by \cite{bvacless}
\begin{equation}\label{current}
j^{\mu}=\epsilon^{\mu\nu}\partial_{\nu}W(\phi(x)),
\end{equation} 
where $\epsilon^{\mu\nu}$ is the antisymmetric symbol in two dimensions with $\epsilon^{01}=1$. Associated with this current (\ref{current}) we have a topological charge given by
\begin{equation}\label{charge}
Q=\int_{-\infty}^{\infty}j^0dx= W(\phi(\infty))-W(\phi(-\infty)).
\end{equation}
Despite the similarities in the values of the charge (\ref{charge}) and the BPS energy (\ref{Ebps}), they have fundamental differences. While the BPS energy is associated with a continuous symmetry and can be identified from the Noether Theorem, the charge (\ref{charge}) is associated with the topology of each solution, and results from the transition of the topological solution in between two minima of the potential.


\subsection{The deformation procedure}

Once we know the behavior of a given model, with its characteristics and general behavior, it is interesting to look for new well-behaved models. In this sense, the Deformation Method \cite{blm} is a powerful tool to find new models in Field Theory. The method consists of choosing a theory with the Lagrangian ${\cal L}(\phi,\partial_\mu\phi)$ having the form (\ref{ori}) and then perform a transformation of the type
\begin{equation}\label{phi_to_chi}
\phi\rightarrow f(\chi).
\end{equation}
We then get
\be\label{14}
{\cal L}(\phi,\partial_{\mu}\phi)=f_\chi^2\;{\cal L}(\chi,\partial_{\mu}\chi)
\ee
where
\begin{equation}\label{defmod}
{\mathcal{L}}(\chi,\partial_{\mu}\chi)=\frac{1}{2}\partial_{\mu}\chi\partial^{\mu}\chi-U(\chi),
\end{equation}
with the potential $U(\chi)$ given by
\begin{equation}\label{newpot}
U(\chi)=\frac{V\left(\phi\rightarrow f(\chi)\right)}{f_{\chi}^2}.
\end{equation}
This is a field redefinition, but if we consider the model \eqref{defmod} described by the potential (\ref{newpot}) as a new model, and call it the deformed model, in this case there is a $\bar{W}$ function such that the first-order differential equations
\begin{equation}\label{eq1}
\frac{d\chi}{dx}= {\bar{W}}_{\chi},
\end{equation}
with $\bar{W}_\chi=W_\phi(\phi\to f(\chi))/f_\chi$, provide solutions for the field $\chi(x)$.  In particular, among the characteristics of the new model, we can highlight that the total energy of the solution is
\begin{equation}\label{Ebpsdeformed}
{E}=|\bar{W}\left(\chi (\infty)\right)-\bar{W}\left(\chi(-\infty)\right)|,
\end{equation}
since the deformed model also has a first order structure. However, the deformed model engenders another important property: the transformation (\ref{phi_to_chi}) allows us to find the solution for the field $\chi(x)$ without the need to deal directly with equation (\ref{eq1}). What we have is that the solution to the new field is obtained by inverting the transformation (\ref{phi_to_chi}). Thus, one finds
\be\label{sol}
\chi(x)=f^{-1}(\phi(x)),
\ee
where $\phi(x)$ represents a solution for the previous model. In this sense, the equation (\ref{sol}) represents the main link between the $\phi$-model and the deformed $\chi$-model. For more details, see \cite{blm}.

\section{Model}
\begin{figure}[t]
\centerline{\includegraphics[height=14em]{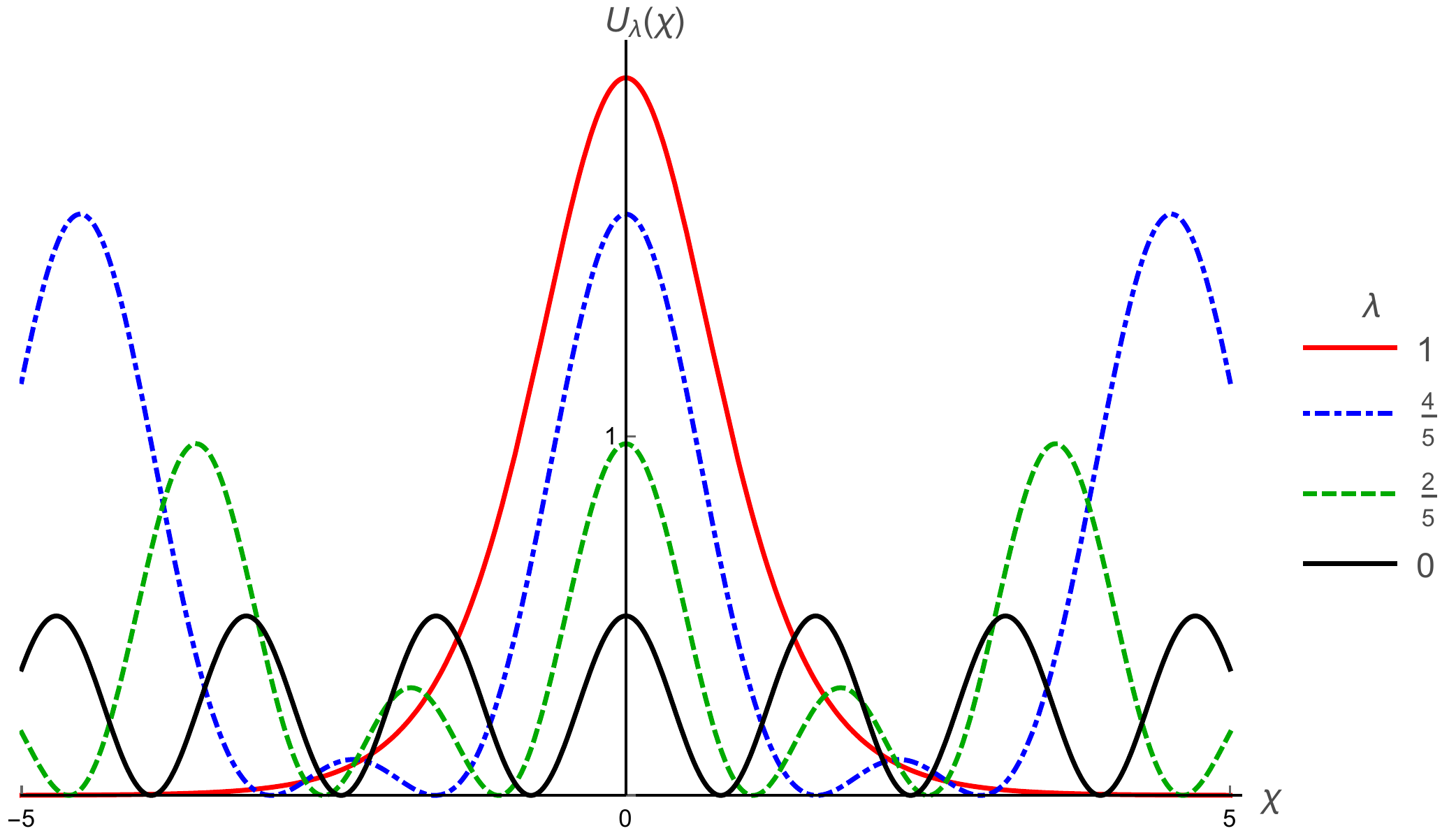}}
\caption{The potential (\ref{pot1}) for some values of $\lambda$ and $\theta^2=1-\lambda$.
As $\lambda$ increases in the interval $[0,1]$, the model evolves from the sine-Gordon model at $\lambda=0$ to a double sine-Gordon model and finally to the vacuumless model, at $\lambda=1$.}\label{fig1}
\end{figure}

In this work we follow the deformation procedure and introduce a new model generated by the deformation function
\begin{equation}\label{d1}
f_\lambda(\chi)=\tanh\left(\frac{1}{\theta\sqrt{2-\theta^2}}\text{tanh}^{-1}\left(\theta\frac{\text{sc}\left(\chi,\lambda\right)}{\sqrt{2-\theta^2}}\right)\right)
\end{equation}
which is applied to the $\phi^4$ model (\ref{phi4}).  Here $\theta$ is a real parameter $(\neq\pm\sqrt{2})$ and  the function $\text{sc}^{-1}(\chi,\lambda)$ is one of the Jacobi Elliptic Functions, defined by the ratio
\begin{equation}
\text{sc}(\chi,\lambda)=\frac{\text{sn}(\chi,\lambda)}{\text{cn}(\chi,\lambda)}, 
\end{equation}
where $\text{sn}(\chi,\lambda)$ and $\text{cn}(\chi,\lambda)$ are the Jacobi elliptic sine and cosine, respectively. Here we have 
\bes\label{jacobirelations}\ben
\text{cn}^2(\chi, \lambda)+\text{sn}^2(\chi, \lambda)&=&1\\
\text{dn}^2(\chi, \lambda)+\lambda\text{sn}^2(\chi, \lambda)&=&1
\een\ees
where $\lambda$ is a parameter in the interval  $[0,1]$. In particular for $\lambda=0$ we have $\text{sn}(\chi,0)=\sin(\chi)$, $\text{cn}(\chi,0)=\cos(\chi)$ and  $\text{dn}(\chi,0)=1$, where we retrieve the basic trigonometric relations, and  for $\lambda=1$ we have $\text{sn}(\chi, 1)=\text{tanh}(\chi)$ and $\text{cn}(\chi, 1)=\text{dn}(\chi, 1)=\text{sech}(\chi)$, which lead us to the hyperbolic functions. 
The new model found from the deformation function (\ref{d1}) and the potential \eqref{phi4} is
\begin{equation}\label{pot1}
\text{U}(\chi, \lambda)=\frac{1}{2} \frac{\left(2\text{cn}^2 (\chi, \lambda)-\theta ^2\right)^2}{\text{dn}^2(\chi, \lambda)}.
\end{equation}
Its behavior is depicted in Fig.~\ref{fig1}. The potential (\ref{pot1}) has $\mathbb{Z}_2-$symmetry and is invariant under transformation $\phi\to\phi+2K_\lambda$, where $K_\lambda=\text{cn}^{-1}(0,\lambda)$.  In this case we find $W_\chi=\pm (2\text{cn}^2(\chi,\lambda)-\theta ^2)/\text{dn}(\chi, \lambda)$, which implies that the $W$ function of the model is
{\small \begin{eqnarray}
\nonumber\text{W}(\chi,\lambda)&=&\frac{2}{\lambda}\left(\text{am}(\chi,\lambda)-\sqrt{1-\lambda}\text{tan}^{-1}(\sqrt{1-\lambda}\text{sc}\left(\chi,\lambda\right))\right)\\
&~&-\frac{\theta^2\cos^{-1}(\text{cd}(\chi,\lambda)\sqrt{1-\text{cd}^2(\chi,\lambda)}\text{dn}(\chi,\lambda)}{(1-\lambda)\text{sn}(\chi,\lambda)},
\end{eqnarray}}where $\text{am}(\chi, \lambda)=d\left(\text{dn}(\chi, \lambda)\right)/d\chi$ is the {\it Jacobi amplitude}. Although we initially present the model with two parameters, we are interested in situations where only one parameter is necessary to describe the changes in the model. So  we assume that $\theta=\theta(\lambda)$. We also impose on $\theta(\lambda)$ the conditions $\theta (0)=1$ and $\theta(1)=0$ in order to find the vacuumless model \cite{bvacless} for $\lambda=1$ and the sine-Gordon model \cite{caudrey} for $\lambda=0$. As a consequence we obtain the particular cases $W(\chi,1)=4\tan^{-1}(e^{\chi})-\pi$ and $W(\chi,0)=\sin (2\chi)/2$, so the sine-Gordon and vacumless models appear in the system as
\bes\ben\label{sgpot}
U(\chi, 0)&=&\frac{1}{2} \cos ^2(2 \chi ),\\
U(\chi,1)&=&2\, \text{sech}^2(\chi ).\label{vlpot}
\een\ees
\begin{figure}[t]
\centerline{\includegraphics[height=14em]{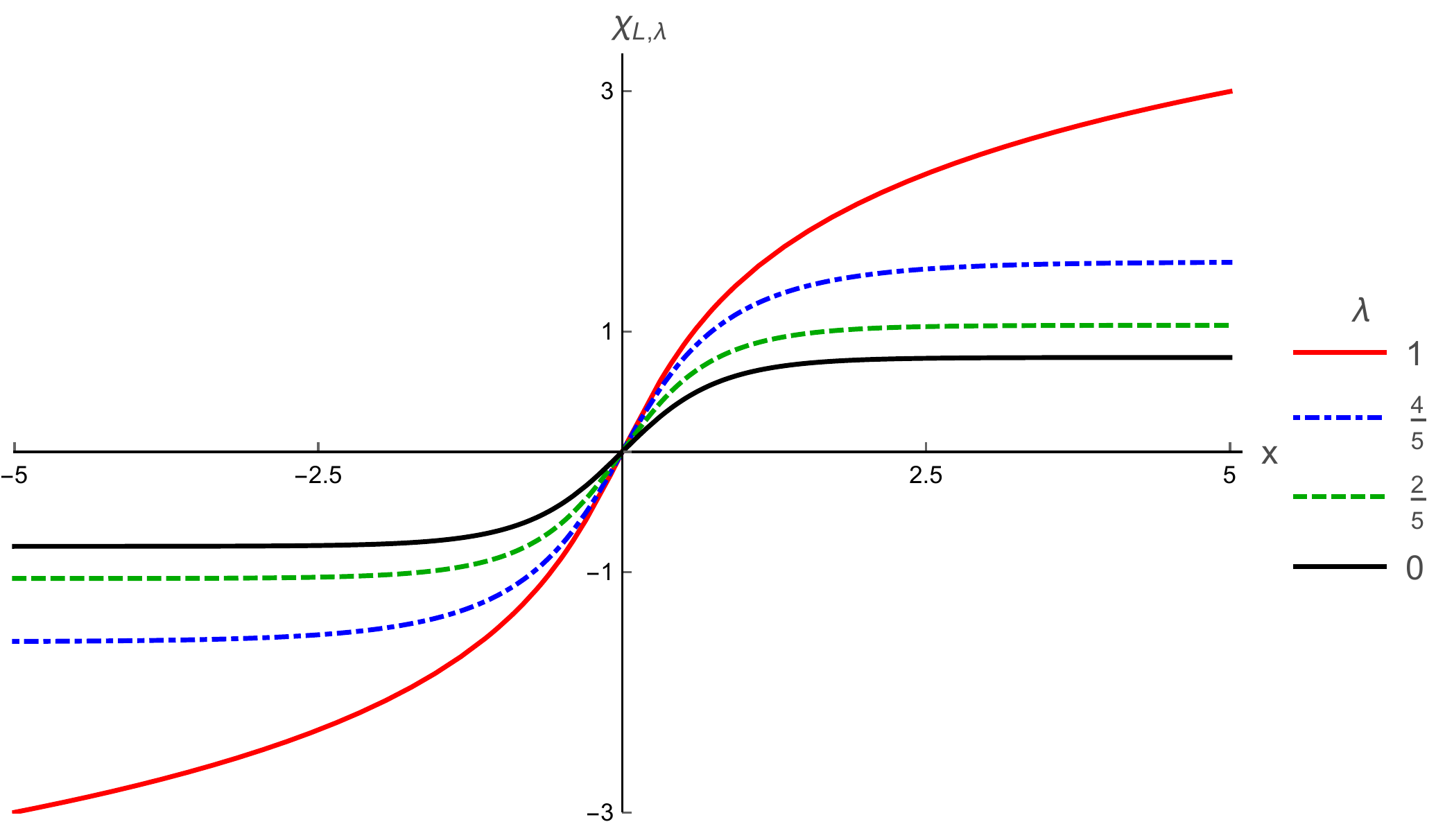}}
\caption{The solution $\chi_{L,\lambda}$ for some values of $\lambda$ and $\theta^2=1-\lambda$. Here we observe how the large kink of the model evolves from the sine Gordon kink to the vacuumless solution.}\label{fig2}
\end{figure}
\begin{figure}[t]
\centerline{\includegraphics[height=14em]{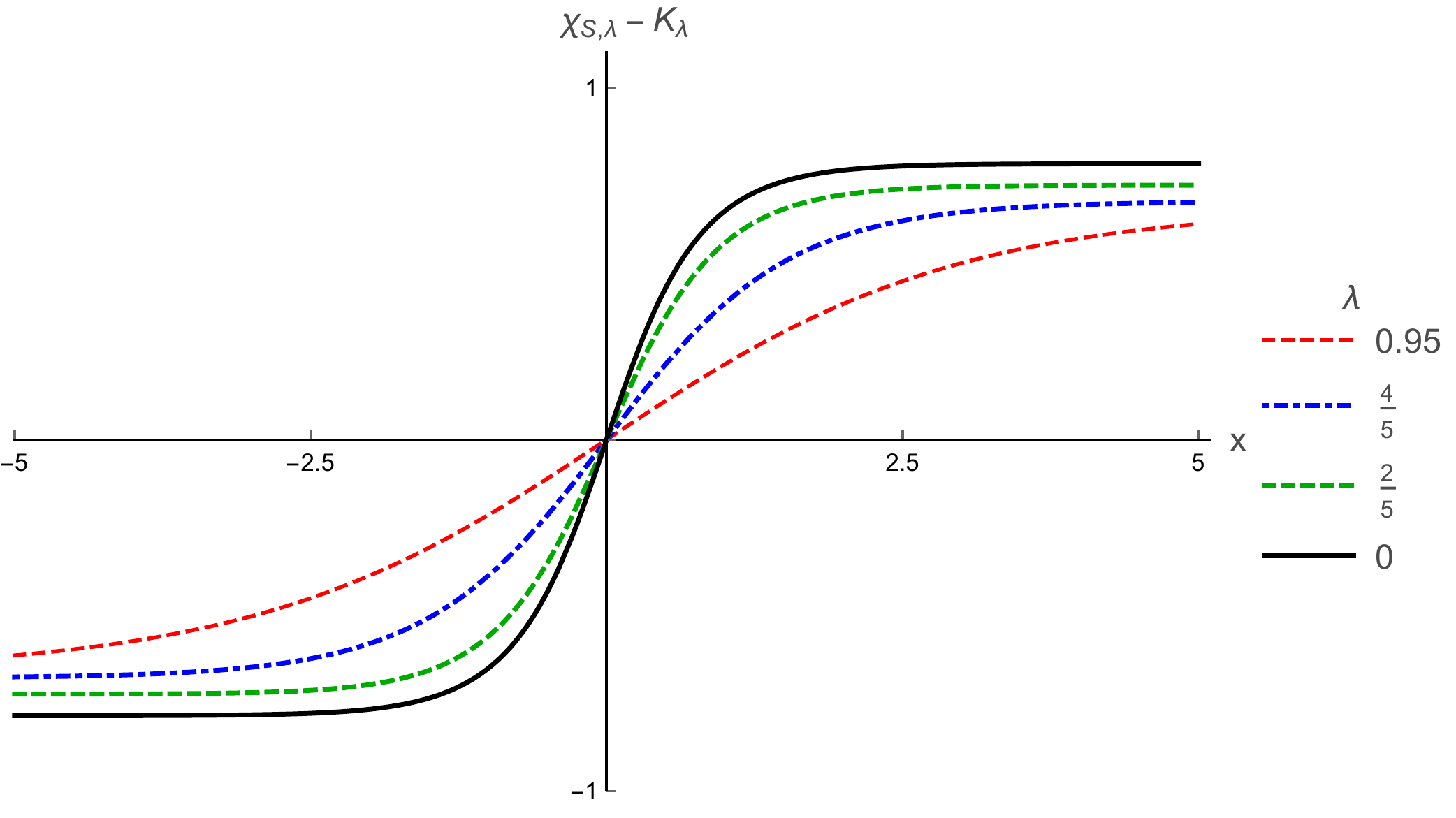}}
\caption{The solution $\chi_{S,\lambda}-K_\lambda$ for some values of $\lambda$ and
$\theta^2=1-\lambda$. Here we observe how the small kink tend to disappear in the limit $\lambda\to1$.}\label{fig3}
\end{figure}
\begin{figure}[t]
\centerline{\includegraphics[height=14em]{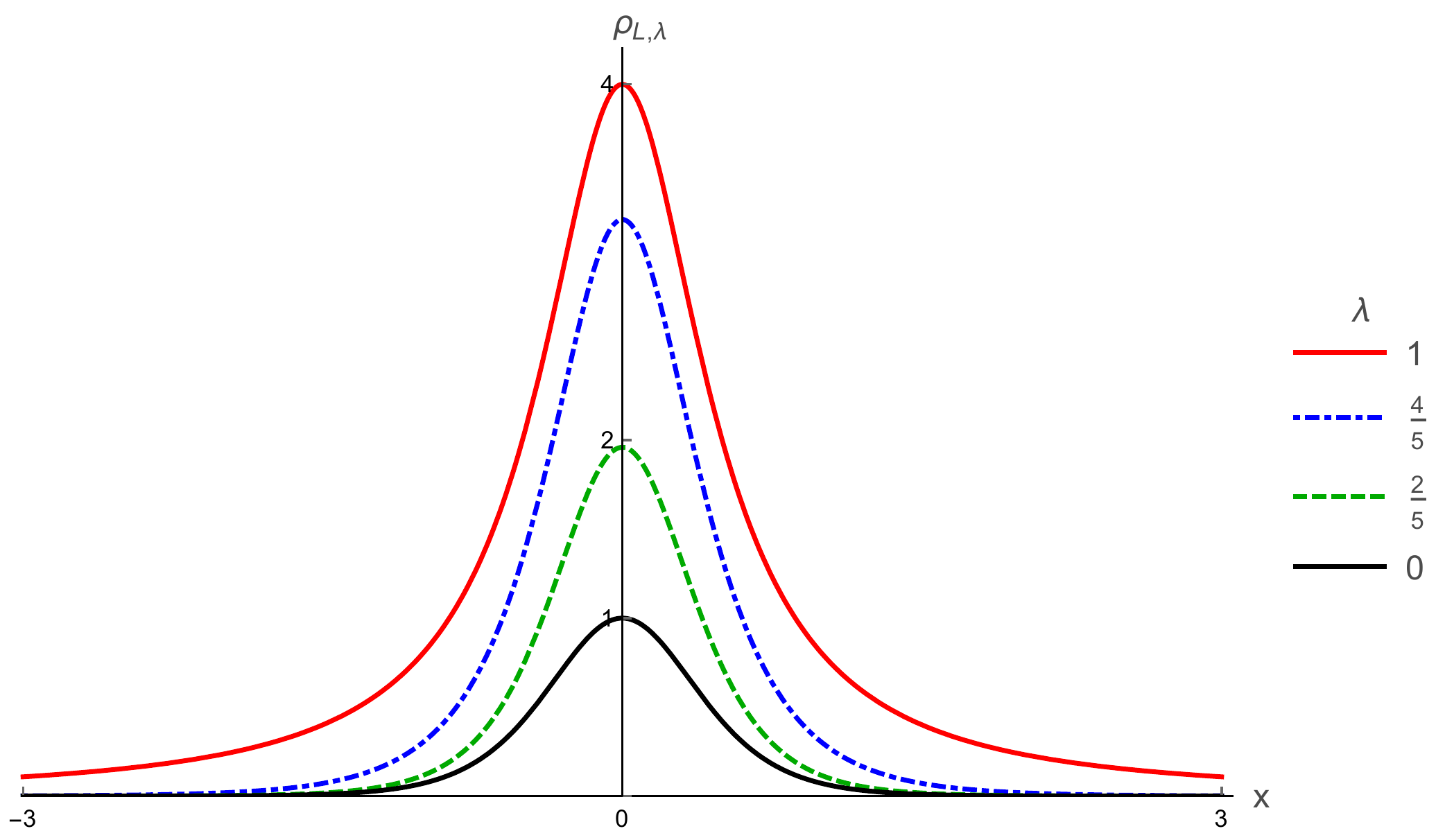}}
\caption{The energy density (\ref{rhoL}) for some values of $\lambda$. Here one notes how
$\rho_{L,\lambda}$ behaves as $\lambda\to 1$, becoming more diffuse but keeping its localized profile.}\label{fig4}
\end{figure}

The parameter $\lambda$ has an interesting behavior: when it increases from $0$ to unity, the model changes from the sine-Gordon to the vacuumless model. Physically, it transforms the periodic sine-Gordon potential \eqref{sgpot} to a non-periodic one, the hyperbolic potential \eqref{vlpot} which defines the vacuumless model. The distinct solutions appear in Fig.~\ref{fig2} and in Fig.~\ref{fig3} one illustrates how the energy density becomes more and more diffuse, as $\lambda$ increases from zero to unity.

Once the model has been presented, we must solve the first order equation
\begin{equation}\label{1eq}
\chi'= \frac{2\text{cn}(\chi, \lambda)^2-\theta ^2}{\text{dn}(\chi, \lambda)}.
\end{equation}
Equation (\ref{1eq}) has two infinite set of solutions. The first one we call {\it large kinks} and are given by 
{\small \begin{equation}\label{largek}
\chi_{L,\lambda}(x)=\text{sc}^{-1}\left(\frac{\sqrt{2-\theta ^2}}{\theta} \tanh\left(\theta\sqrt{2-\theta ^2}x\right),\lambda\right)\!+\!
2n\text{K}_{\lambda}.
\end{equation}}
The behavior of the solution (\ref{largek}) is depicted in Fig.~\ref{fig2} for $n=0$. It asymptotically approaches $\chi_{L,\lambda}(\pm\infty)=\pm\text{sc}^{-1}(\sqrt{2-\theta ^2}/\theta,\lambda)$ and in the vicinity of the origin behaves like $\chi_{L,\lambda}(x\simeq 0)\simeq (2-\theta^2)x +\mathcal{O}(x^2)$. We note that as $\theta\rightarrow 0$,  $\chi_{L,\lambda}(\pm\infty)$ tends to diverge and $\chi_{L,\lambda}(x\simeq 0)$ remains well-behaved. For $\theta\to 1$ we do not have pathologies in the solution.  Thus, the large kink describes, as a function of the $\lambda$  parameter, systems that transit between the sine-Gordon kink and the vacuumless solution. Particularly, we have
\bes\ben
\chi_{L,1} (x)&\!=\!&\text{sinh}^{-1}(2 x) \text{{\small (vacuumless solution)}};\\
\chi_{L,0} (x)&\!=\!&\text{tan}^{-1}\!\left(\tanh(x)\right)\text{{\small (sine-Gordon kink)}}.
\een\ees

The second set of solutions we find from equation (\ref{1eq}) is
{ \begin{equation}\label{smallk}
\chi_{S,\lambda}(x)=\text{sc}^{-1}\left(\frac{\theta  \tanh \left(\theta  \sqrt{2-\theta ^2} x\right)}{\sqrt{\left(2-\theta ^2\right) (1-\lambda)}}, k\right)\!+\!(2n+1)\text{K}_{\lambda},
\end{equation}}\\
and we call it {\it small kink}.  Its shape is depicted in Fig.~\ref{fig3} for $n=0$, except for the phase $K_\lambda$. The solution (\ref{smallk}) asymptotically approaches $\chi_{S,\lambda}(\pm\infty)=\pm\text{sc}^{-1}(\theta/\sqrt{(2-\theta ^2) (1-\lambda)}, \lambda)+K_{\lambda}$ and in the neighborhoods of $x=0$ behaves as  $\chi_{S,\lambda}(x\simeq 0)\simeq K_\lambda+\theta^2x/\sqrt{1-\lambda} +\mathcal{O}(x^2)$. The small kink  has a phase $K_\lambda$, which causes its topological sector to move away from the center of the potential  (\ref{pot1}) to infinity as $\lambda\to 1$. 
With a suitable choice for $\theta(\lambda)$ we can destroy the topological sectors associated with small kinks when $\lambda=1$. Thus, only the topological sector of the vacuumless solution remains at that point. If we drops out the phase in the small kink (\ref{smallk}), it is possible to note that the zero energy solutions also remains, but at infinity.  A simple choice in this direction that also obeys the conditions
$\left(\theta(0),\theta(1)\right)=(1,0)$ is $\theta(\lambda)=\sqrt{1-\lambda}$. With this choice for $\theta(\lambda)$ we can explicitly rewrite the solutions of large and small kinks as 
\bes\label{sol1}\ben
\chi_{L, \lambda}(x)&\!=\!&\text{sc}^{-1}\!\left(\sqrt{\frac{1+\lambda}{1-\lambda}} \tanh (\sqrt{1-\!\lambda ^2} x),\lambda \right)\\
\chi_{S, \lambda}(x)&\!=\!&\text{sc}^{-1}\!\left(\frac{\tanh \left(\sqrt{1-\!\lambda^2} x\right)}{\sqrt{1+\lambda}},\lambda\right)+K_\lambda
\een\ees
So now we have a double sine-Gordon-like model with two field solutions coming from the two manifest topological sectors. Both kinks retrieve the sine-Gordon solution when $\lambda = 0$, but as $\lambda$ grows, such solutions have distinct properties. 
Large kink becomes more diffuse until reaching the solution of the vacuumless model, which has divergent amplitude. It implies that at this point all topological sectors of the model are sent to infinity, except for the sector that is at the center of the potential (\ref{pot1}).  Small kink also becomes more diffuse as $\lambda\to 1$, but is destroyed when $\lambda=1$. As we shall see later, it happens because as the associated topological sector moves away from the center of the potential its energy approaches zero. Moreover, one can show that the mass of the meson in the minima of the potential is given by $m_\lambda^2=U''(\chi=\chi_{min})=4(1-\lambda^2)$, where $\chi_{min}$ is a minimum of (\ref{pot1}). Here, we choose to write the model in terms of that quantity, whenever possible.

We can now perform the analysis of the energy densities of the model. The energy density of the large kink (\ref{largek}) is given by
\begin{equation}\label{rhoL}
\rho_{L,\lambda}(x)=\frac{(1-\lambda ) (1+\lambda)^2 \text{sech}^2\left(\frac{m_\lambda}{2} x\right)}{\left(\cosh \left(m_\lambda x\right)-\lambda\right) \left((1+\lambda )\tanh ^2\left(
\frac{m_\lambda}{2} x\right)+1\right)}
\end{equation}
and its shape is depicted in Fig.~\ref{fig4}. It makes the transition between the curves $\rho_{L,1}(x)=4/(1+4 x^2)$ and $\rho_{L,0}(x)=\text{sech}^2(2 x)$. Asymptotically equation (\ref{rhoL}) decay as  $\rho_{L,\lambda}(x\to\infty)\simeq 8(1-\lambda)(1+\lambda)^2e^{-2 m_\lambda x}/(2+\lambda)+\mathcal{O}(e^{-4 m_\lambda x})$, where we found the particular case $\rho_{L,0}(x\to\infty)\simeq 4e^{-4x}$. We don't have information for the case $\lambda=1$ in the general asymptotic expansion for the energy density, but a direct approach in $\rho_{L,1}(x)$ lead us to the expression $\rho_{L,1}(x\to\infty)\simeq 1/x^2 +\mathcal{O}(x^{-4})$, which decays much slower than the exponential. This change in the asymptotic behavior is due to the mass scale of the quantum meson which, for $\lambda=1$, is zero. On the other hand the behavior of (\ref{rhoL}) in the vicinity of $x=0$  is given by $\rho_{L,\lambda}(x\simeq 0)\simeq(1+\lambda)^2-(1+\lambda)^3 (4-\lambda ^2-\lambda) x^2+\mathcal{O}(x^3)$, which shows that the central portions of the energy density increases and becomes a bit more concentrated as $\lambda\to 1$.

\begin{figure}[t]
\centerline{\includegraphics[height=14em]{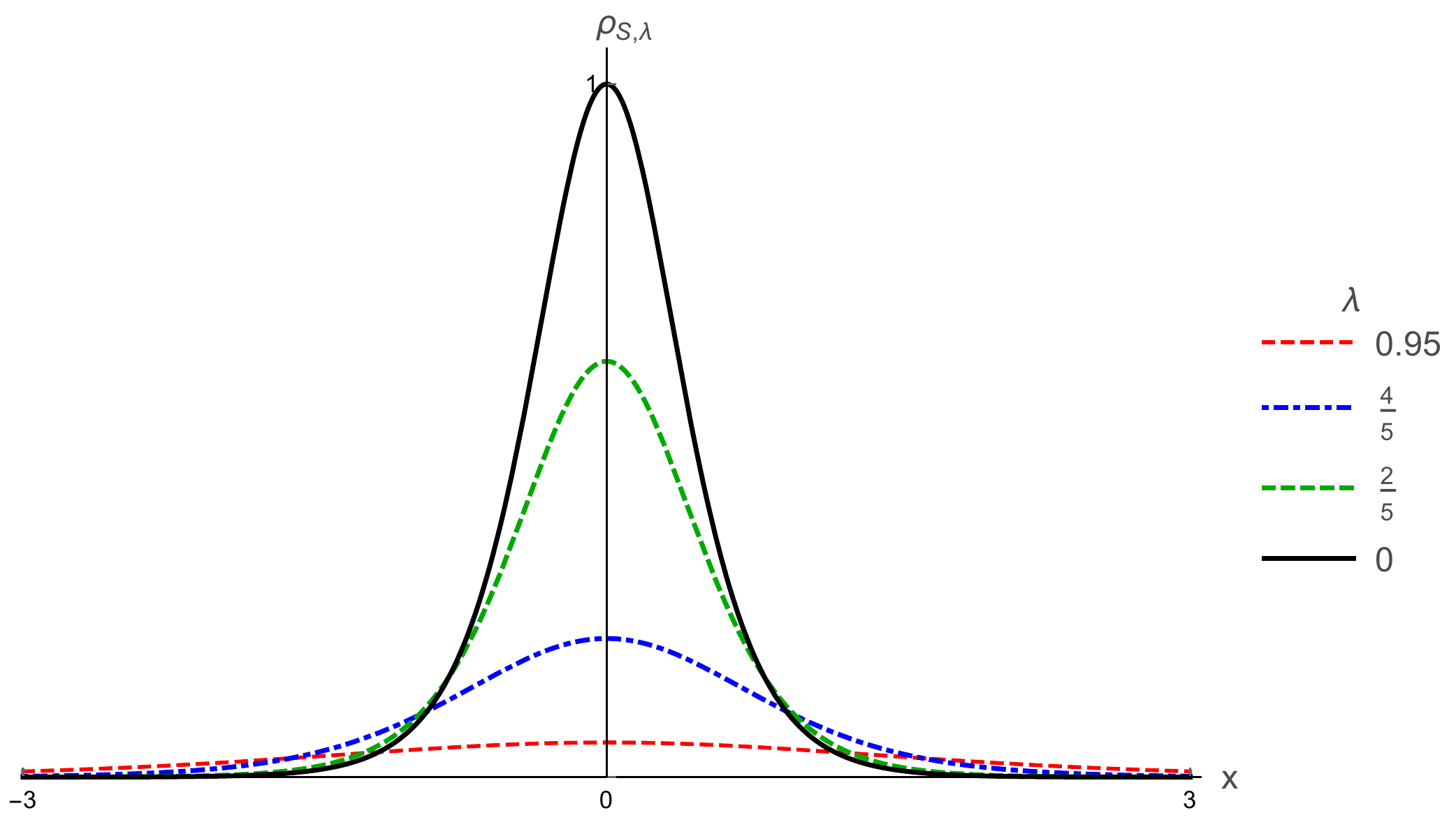}}
\caption{The energy density (\ref{rhoS}) for some values of $\lambda$. Here one notes how $\rho_{S,\lambda}$ behaves as $\lambda\to 1$, becoming more delocalized, disappearing at $\lambda=1$.
}\label{fig5}
\end{figure}

For the small kink we have the energy density given by
\begin{equation}\label{rhoS}
\rho_{S,\lambda}(x)=\frac{(1-\lambda ) (1+\lambda )^2 \text{sech}^2\left(\frac{m_\lambda}{2} x\right)}{\left(\cosh \left(m_\lambda x\right)\!+\!\lambda\right) \left(\tanh ^2\left(\frac{m_\lambda}{2}x\right)+1+\lambda\right)}
\end{equation}
and its behavior is depicted in Fig. \ref{fig5}.  Equation (\ref{rhoS}) transits between the curves $\rho_{S,1}(x)=0$ and $\rho_{L,0}(x)=\text{sech}^2(2 x)$.  Asymptotically it decay as  $\rho_{S,\lambda}(x\to\infty)\simeq 8(1-\lambda)(1+\lambda)^2e^{-2 m_\lambda x}/(2+\lambda)+\mathcal{O}(e^{-4 m_\lambda x})$, revealing that we can not distinguish the energy densities of the fields (\ref{largek}) and (\ref{smallk}) when $x\to\infty$. For $x\simeq 0$ we have $\rho_{S,\lambda}(x\simeq 0)\simeq(1-\lambda )-(1-\lambda )^2 (\lambda +4) x^2+\mathcal{O}(x^3)$. It shows that, despite the similar asymptotic behavior (for $\lambda\neq1$), the evolution of (\ref{rhoL}) and (\ref{rhoS}) in terms of the parameter $\lambda$ is very different.  The height and width of the energy density (\ref{rhoL}) grow with $\lambda$, becoming more diffuse but still with a localized profile. Meanwhile, as $\lambda$ grows, the width of the energy density (\ref{rhoS}) increases, but its central portion decreases and becomes less concentrated, which implies that as the solution evolves it becomes more diffuse, but also delocalized. As a consequence, the area under the energy density (\ref{rhoS}) become smaller until it disappears, at $\lambda=1$.

\begin{figure}[t]
\centerline{\includegraphics[height=12em]{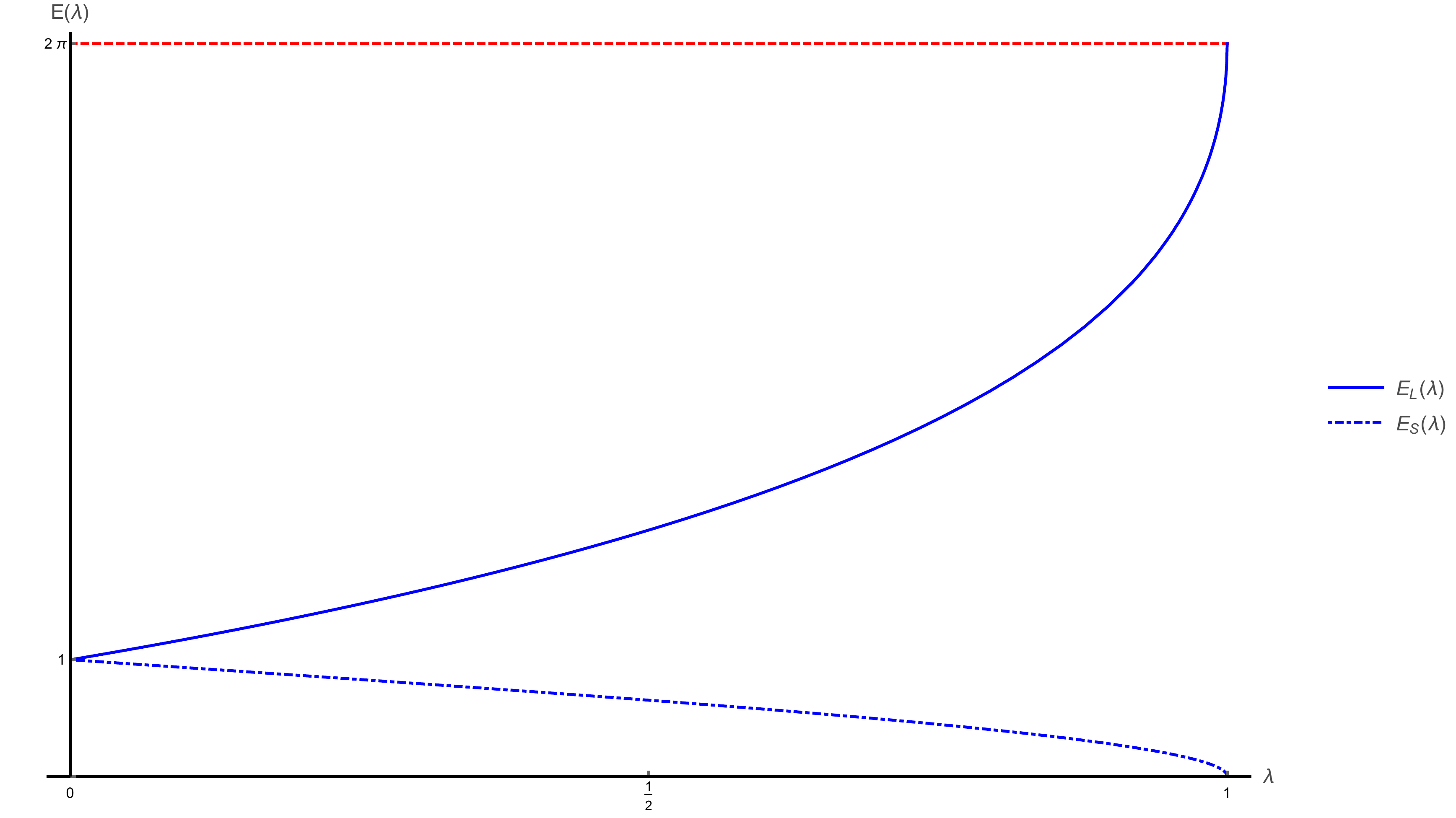}}
\caption{The energy behavior as a functon of $\lambda$. Here we observe that $E_L(\lambda)$ evolves from $E_L(0)=1$ to $E_L(1)=2\pi$ and $E_S(\lambda)$ decreases from $E_S(0)=1$ to $E_S(1)=0$.}\label{fig6}
\end{figure}

Integrating (\ref{rhoL}) and (\ref{rhoS}) we find the energy of large and small kinks, which are given by
\begin{eqnarray}\label{EL}
\nonumber E_{L}(\lambda)&=&\frac{2}{\lambda}\biggl(2 \tan^{-1}\biggl(\sqrt{\frac{1+\lambda}{1-\lambda}}\biggr)-\\
&~&-\sqrt{1-\lambda}\left(2+\lambda\right) \tan ^{-1}\left(\sqrt{1+\lambda}\right)\biggr)
\end{eqnarray}
and 
\begin{eqnarray}\label{ES}
\nonumber E_{S}(\lambda)&=&-\frac{2}{\lambda}\biggl(2 \cot^{-1}\biggl(\sqrt{\frac{1+\lambda}{1-\lambda}}\biggr)-\\
&~&-\sqrt{1-\lambda}\left(2+\lambda\right) \cot^{-1}\left(\sqrt{1+\lambda}\right)\biggr),
\end{eqnarray}
respectively. The corresponding behaviors are depicted in Fig.~\ref{fig6}. Here one observes that the energy of the solutions (\ref{largek}) and (\ref{smallk}) are bounded. The expression (\ref{EL}) is a monotonically increasing function of the $\lambda$ parameter confined in the interval $[1, 2\pi]$, where $E_L(0)=1$ and $E_L(1)=2\pi$, and the expression (\ref{ES}) is a monotonically decreasing function of the $\lambda$ parameter confined in the interval $[0,1]$, with $E_S(0)=1$ and $E_S(1)=0$. Moreover, it is easy to show that (\ref{EL}) and (\ref{ES}) are related by $E_{L}(\lambda)=E_{S}(\lambda)+\frac{\pi}{\lambda}\left(2-(2+\lambda)\sqrt{1-\lambda}\right)$.

\subsection{Linear stability}

\begin{figure}[t]
\centerline{\includegraphics[height=14em]{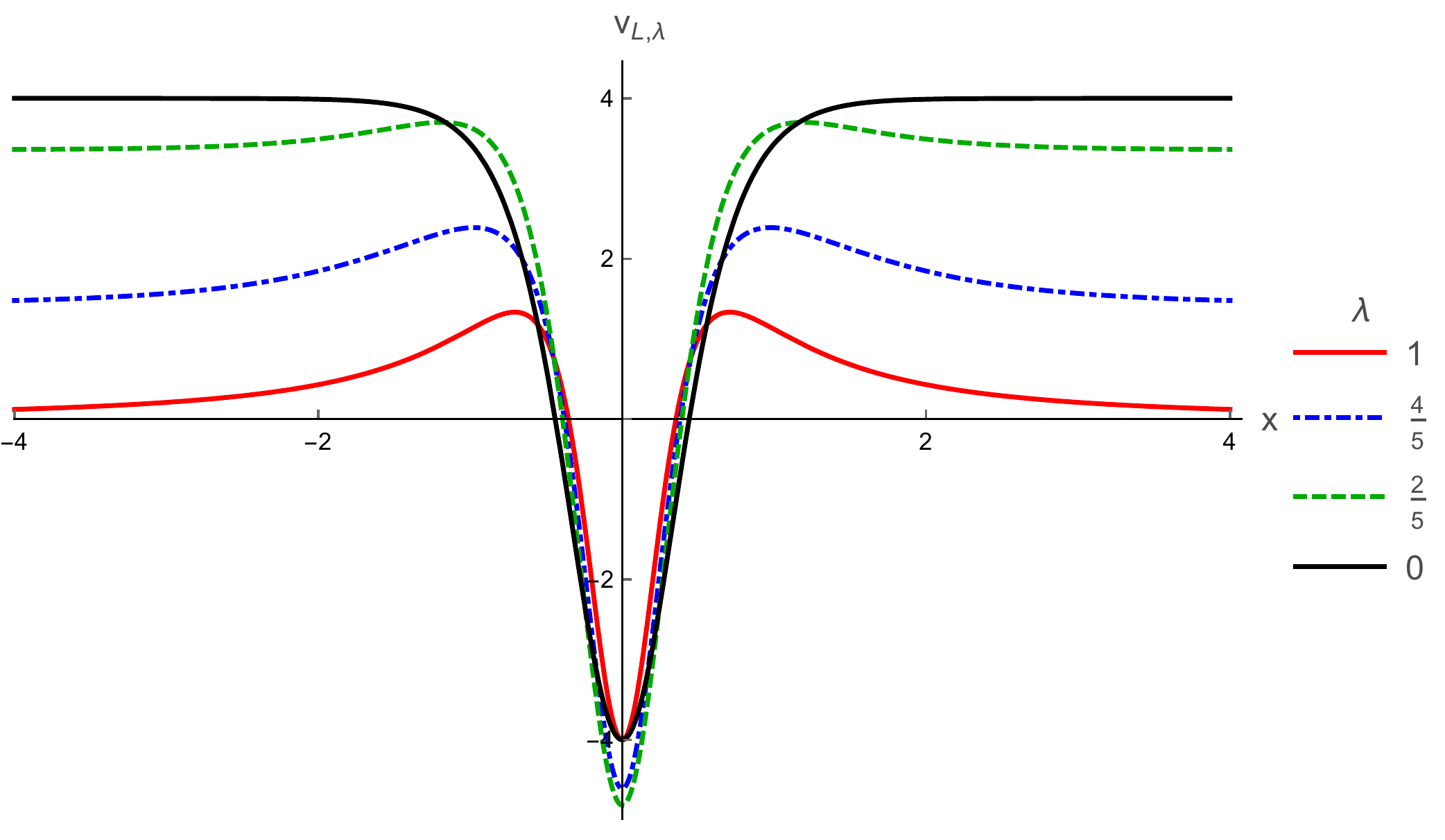}}
\caption{The stability potential $v_{L,\lambda}(x)$ given by (\ref{vL}) for some values of $\lambda$. When $\lambda=0$, we have a reflectionless potential and as $\lambda$ increases the shape of the potential changes to become of the volcano type.}\label{fig7}
\end{figure}

In this section we analyze the stability of the solutions of the models presented so far. The usual procedure is to take a time-dependent perturbation around the static solution written in the form $\chi(x,t)=\chi(x)+\sum_n \eta_n(x)\cos (\omega_n t)$, for small $\eta_n(x)$, and then insert $\chi(x,t)$ into (\ref{equationofmotion}). The procedure gives
\begin{equation}\label{stabilityequation}
\left(-\frac{d^2}{dx^2}+v(x)\right)\eta_n(x)=\omega_n^2\eta_n(x),
\end{equation}
which is a Schr\"odinger-like  equation with a stability potential given by
\bes\label{stabilitypotential}\ben
v(x)&=&\frac{d^2U}{d\chi^2}\biggr{|}_{\chi=\chi(x)}\\
&=&\bar{W}_{\chi\chi}^2\bigr{|}_{\chi=\chi(x)}+\bar{W}_{\chi\chi\chi}\bar{W}_\chi\bigr{|}_{\chi=\chi(x)}.
\een\ees
Inserting the large kink solution (\ref{largek}) into (\ref{stabilitypotential}) we find
\begin{widetext}
\begin{eqnarray}\label{vL}
\nonumber v_{L,\lambda}(x)&=&\frac{1-\lambda^2}{(\cosh\left(m_\lambda x\right)-\lambda)^2 ((1+\lambda) \tanh^2\left(\frac{m_\lambda}{2} x\right)+1)^2}\biggl[2 (2+\lambda)^2\left(\cosh\left(2 m_\lambda x\right)-\frac{\lambda(\lambda +7)+8}{2+\lambda}\cosh\left(m_\lambda x\right)\right)
\\
&~&-(1+\lambda)^2 \left((1+\lambda) (\lambda +4) \text{sech}^2\left(\frac{m_\lambda}{2}x\right)-6\lambda \right) \text{sech}^2\left(\frac{m_\lambda}{2}x\right)+\lambda(\lambda(\lambda(\lambda +2)+19)+48)+24\biggr]
\end{eqnarray}
\end{widetext}
which is depicted in Fig.~\ref{fig7}. For $\lambda=1$ we have $v_{L,1}(x)=4 \left(8 x^2-1\right)/\left(4 x^2+1\right)^2$, which is a volcano potential, and for $\lambda=0$ we have $v_{L,0}(x)=4-8 \text{sech}^2(2 x)$, which is a reflectionless  potential. Equation (\ref{vL}) has a global minimum at $v_{L,\lambda}(0)=(1+\lambda) \left(\lambda ^2+\lambda -4\right);$ this minimum increases or decreases, depending on $\lambda$ being above or below the point at $\tilde{\lambda}=\frac{1}{3} \left(\sqrt{13}-2\right)$.  It shows that the potential (\ref{vL}), as $\lambda$ grows, becomes deeper and after $\lambda=\tilde{\lambda}$ it comes back to its initial deepness. 

\begin{figure}[t]
\centerline{\includegraphics[height=14em]{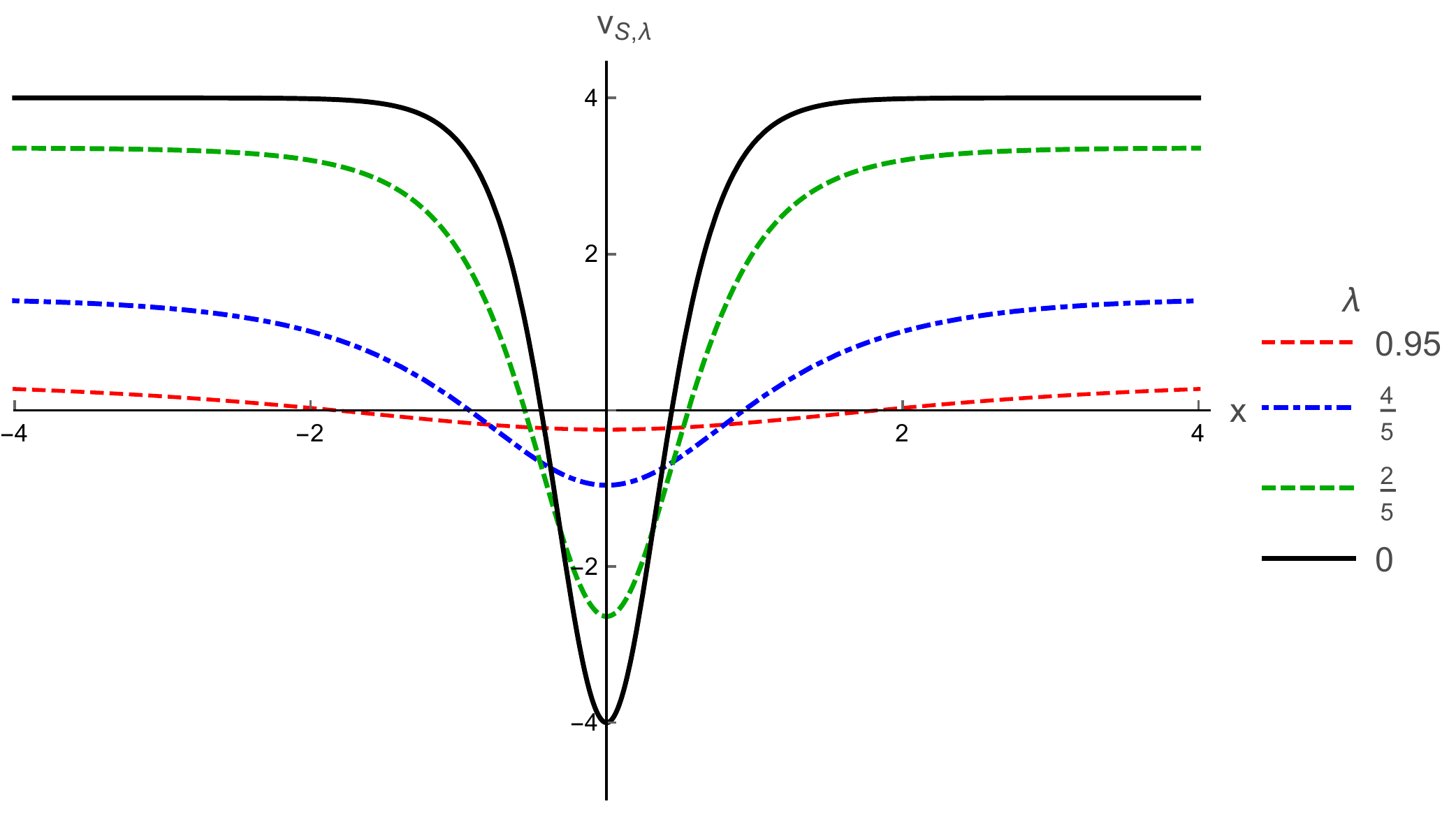}}
\caption{The stability potential $v_{S,\lambda}(x)$ (\ref{vS}) for some values of $\lambda$. Here we observe that as $\lambda$ increases, its depth diminishes and disappears at $\lambda=1$.}\label{fig8}
\end{figure}

The change in the shape of the stability potential, in this case, is due to the behavior of the meson mass in the minima of (\ref{pot1}). As $\lambda\to 1$ we have $m_\lambda^2\to 0$. Note that $v_{L,\lambda}(0)<0$  for all $\lambda$  and  translational invariance requires the existence of at least one bound state, so the transition between the reflectionless and the volcano shapes in this case describes the transition from systems with massive meson to systems having a massless meson. This is another relevant physical behavior induced by the parameter $\lambda$, which will lead to distinct possibilities when used to describe braneworld scenarios, as we discuss in Sec.~\ref{sec.bw}.

Now, inserting the small kink solution (\ref{smallk}) in the general expression for the stability potential (\ref{stabilitypotential}), we find 
\begin{widetext}
\begin{eqnarray}\label{vS}
\nonumber v_{S,\lambda}(x)&=&\frac{1-\lambda^2}{(\cosh\left(m_\lambda x\right)+\lambda)^2 (\tanh^2(\sqrt{1-\lambda ^2}x)+1+\lambda)^2}\biggl[2 (2+\lambda)^2 \left(\cosh\left(2 m_\lambda x\right)-\frac{(1-\lambda) \lambda+8}{2+\lambda}\cosh\left(m_\lambda x\right)\right)\\
&~&-2 \lambda  \left(\lambda  \left(\lambda^2+\lambda -1\right)+3\right) \text{sech}^2\left(\frac{m_\lambda}{2} x\right)-\left(\lambda ^3-5 \lambda +4\right) \text{sech}^4\left(\frac{m_\lambda}{2}x\right)-\lambda ^2 (21-(\lambda -6) \lambda )+24\biggr]
\end{eqnarray}
\end{widetext}
which is depicted in Fig.~\ref{fig8}. For $\lambda=1$ we have $v_{S,1}=0$ and for $\lambda=0$ we have $v_{S,0}=4-8 \text{sech}^2(2 x)$, as expected. At the center we have  $v_{S,\lambda}(0)=(\lambda -1) (\lambda +4)$, showing that the potential (\ref{vS}) becomes shallower as $\lambda$ grows, finally desappearing when $\lambda=1$. Now, as $\lambda$ grows, the stability potential keep its reflectionless shape. As a consequence, the bound states of (\ref{vS}) becomes less expressive as $\lambda\to1$ and disappears when $\lambda=1$.

Asymptotically, both  (\ref{vL}) and (\ref{vS})  approaches $v_{L,\lambda}(\pm\infty)=v_{S,\lambda}(\pm\infty)=m_{\lambda}^2$. Moreover,  one notes that the hamiltonian ${H}=-d^2/dx^2+v(x)$ can be rewritten as ${H}=S^{\dag}S$ with $S^{\dag}=- d/dx-W_{\phi\phi}$ by using the equation (12).  Thus we have $\omega_n^2\geq0$ for all values of $n$, since $H$ is non-negative. As equation (\ref{stabilityequation}) does not admit negative energy modes, the stability of the solution is then ensured. 

\begin{figure}[t]
\centerline{\includegraphics[height=14em]{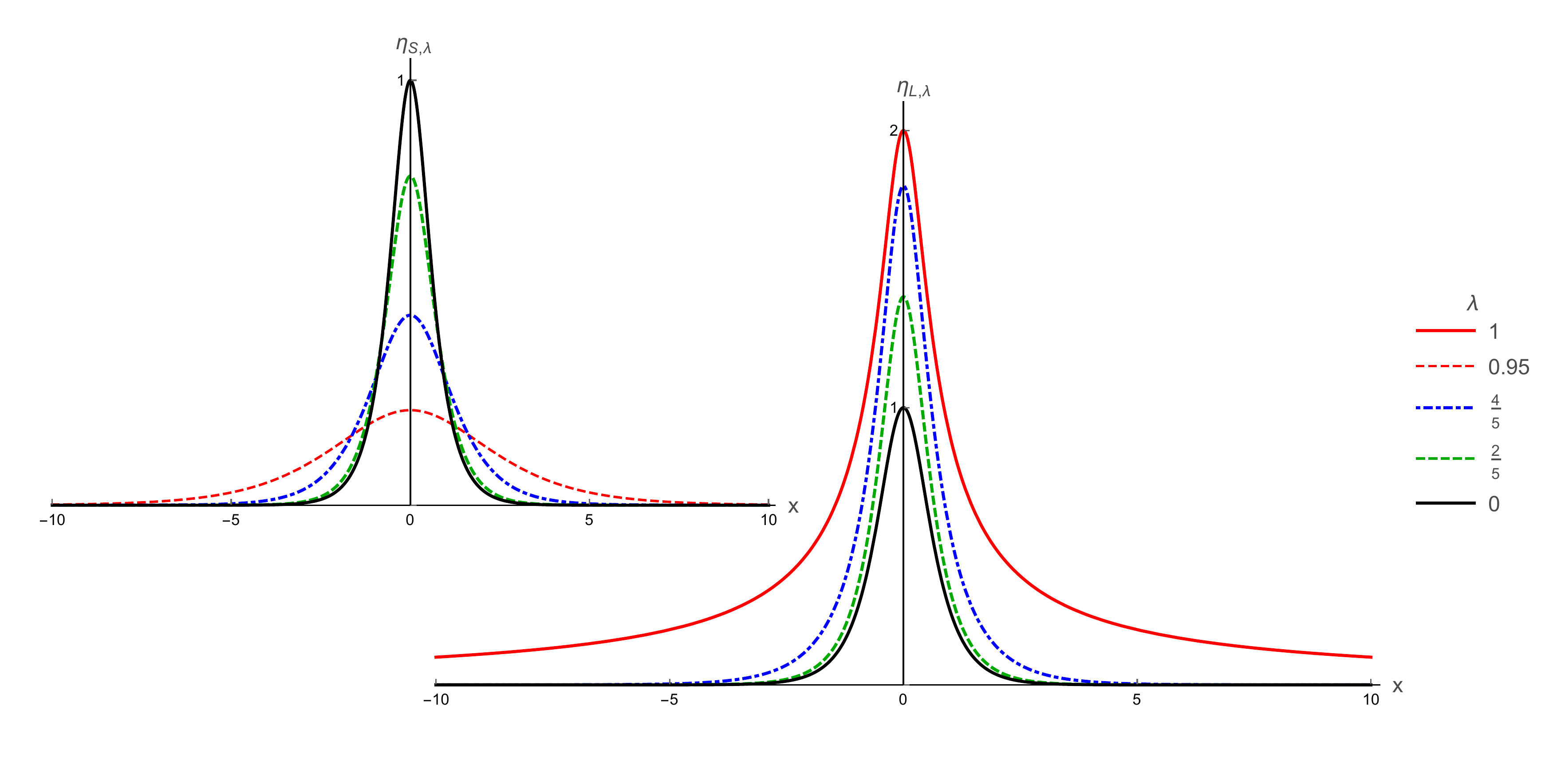}}
\caption{The zero modes for the large kink (\ref{largek}) (right) and the small kink (\ref{smallk}) (left), depicted for some values of $\lambda$. Both modes behave adequately for $\lambda\neq 1$. However, for $\lambda=1$, we have $\eta_{S,\lambda=1}=0$, meaning that the small topological sector disappears, while $\eta_{L,\lambda=1}(x\simeq\infty)\simeq 1/x^2$, which is a consequence of the presence of the massless meson.}\label{fig9}
\end{figure}

The translational invariance of the solutions we presented so far  implies the existence of at least one bound state for each topological sector, which is given by the zero mode of the equation (\ref{stabilityequation}). In formula, if we represent the zero-mode by $\eta_0(x)$, it is the derivative of the field solution,  
\begin{equation}
\eta_0(x)=\frac{d\chi}{dx}.
\end{equation}
For large and small kinks the shapes are depicted in Fig.~\ref{fig9}, where one can observe how (\ref{vL}) holds the nice behavior of its zero-mode, and how the zero mode of (\ref{vS}) gets smaller until it disappears when $\lambda=1$.   

\section{Braneworld}\label{sec.bw}

Models described by scalar fields have direct applications in Gravitation, providing braneworld scenarios for thick branes. In this context, the scalar field acts as a source of gravity around the brane, and thus describes how gravity behaves throughout the bulk. The system we are interested consists of a 3-brane embedded in a (4+1) spacetime with an extra dimension of infinite extent. The background geometry can be written in terms of a static warped metric given by
\begin{equation}\label{metric}
ds^2_5=g_{ab}dx^a dx^b=e^{2A(y)}ds^2_4-dy^2.
\end{equation}
Here $a,  b=0,..., 4$, $\mu, \nu=0,..., 3$, $ds^2_4=\eta_{\mu\nu}dx^{\mu}dx^{\nu}$, and the $y$-coordinate describes the extra spatial dimension. The functions $A(y)$ and  $e^{A(y)}$ are called warp function and warp factor, respectively, and are assumed to depend only on the extra dimension.

In this braneworld scenario, we are interested in models that can be described by the action
\begin{equation}\label{action}
S=\int d^5x\sqrt{|g|}\left(-\frac{1}{4}R+\mathcal{L}\right),
\end{equation}
where $\mathcal{L}(\phi,\partial_{a}\phi)=\frac{1}{2}g_{ab}\partial^{a}\phi\partial^{b}\phi-U(\phi)$  is the Lagrangian for the scalar field and, for simplicity, we assume $4\pi G_5=1$. We also assume that $\phi=\phi(y)$, i.e., the scalar field only depends of the extra dimension. 

The Einstein equations that follows from the action (\ref{action}) are
\begin{equation}\label{ee}
G_{ab}=2T_{ab},
\end{equation}
where $G_{ab}$ is the Einstein tensor and $T_{ab}$ is the energy-momentum tensor, similar to the expression \eqref{emt}. The $00$ and the $44$ components of (\ref{ee}) are given by, respectively
\bes\label{openee}\begin{eqnarray}
 6A'^2&=&\phi'^2-2U,\\
3A''+6A'^2&=&-\phi^{\prime 2}-2 U,
\end{eqnarray}\ees
with the prime denoting derivation in respect to the coordinate $y$. With equations (\ref{openee}) at hands we can subtract the first equation from the second to obtain
\begin{equation}\label{asecondorder}
A''=-\frac{2}{3}\phi'^2. 
\end{equation}
Equation (\ref{asecondorder}) provides a way to rewrite the system in terms of first order equations. For this, we introduce a function $W(\phi(y))$  in the equations through the relation
\begin{equation}\label{afirstorder}
A'=-\frac{2}{3}\,W(\phi(y)).
\end{equation}
As a consequence, the equation providing the solution for the scalar field is now 
\begin{equation}\label{phiprimeequalswphi}
\phi'=W_\phi.
\end{equation}
In order to solve the equations of motion, these two first order equations requires that the potential obeys
\begin{equation}\label{potb}
U(\phi)=\frac12\,W_\phi^2-\frac43\, W^2.
\end{equation}

The above equations \eqref{afirstorder} and \eqref{phiprimeequalswphi} constitute the first order framework and can be used to analyze possible scenarios of thick brane that can be generated by the models presented in the previous section.

\begin{figure}[t]
\centerline{\includegraphics[height=14em]{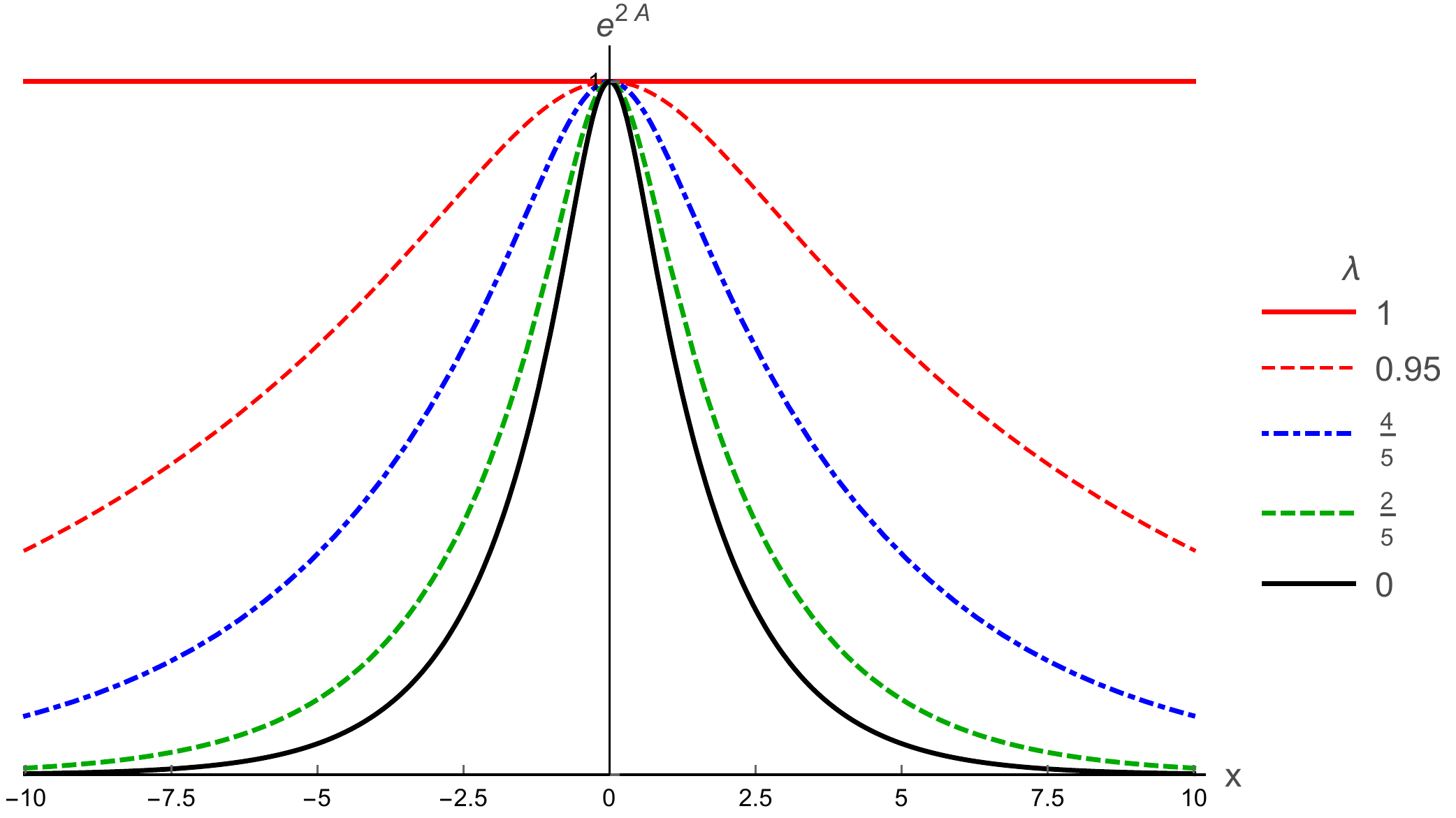}}
\caption{The warp factor that follows from (\ref{warpfunction}) for some values of $\lambda$. Here we see how it evolves from the sine-Gordon case when $\lambda=0$ to a 5-dimensional flat spacetime when
$\lambda=1$.}\label{fig10}
\end{figure}
\begin{figure}[t]
\centerline{\includegraphics[height=14em]{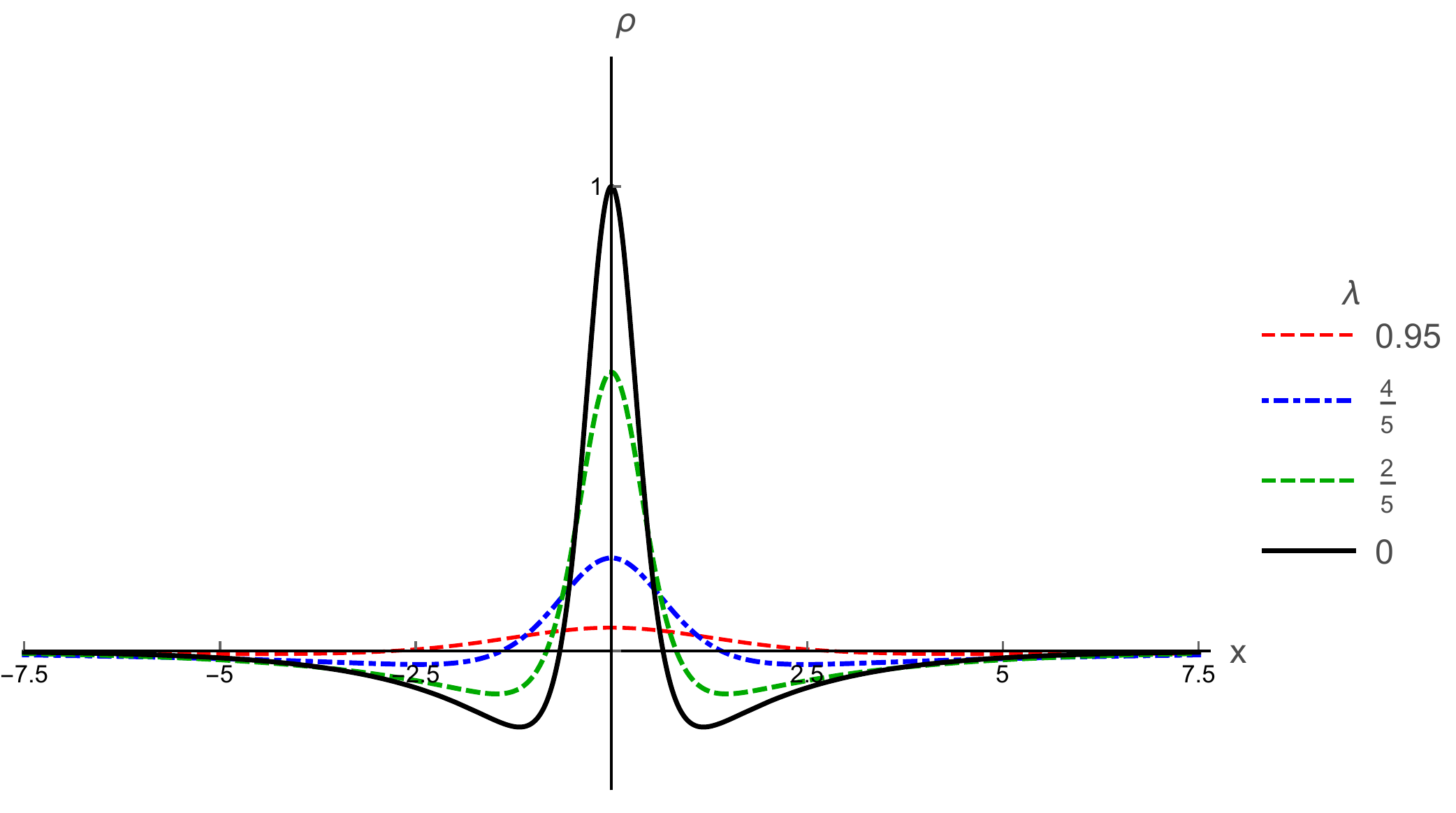}}
\caption{The energy density for some values of $ \lambda$. Here we see how it becomes delocalized as
$\lambda\to 1$. For $\lambda=1$, there is no more energy concentration around the origin and the brane disappears from the background geometry.}\label{fig11}
\end{figure}

The analysis of the thick branes scenario generated by the large kink (\ref{largek}) is similar to the case of small kinks \eqref{smallk}, so we concentrate on the brane generated by the small kink. To simplify the investigation, let us perform a shift in the field, $\phi\to\phi-K_{\lambda}$. In this case, one gets
\ben
W(\phi)&=&\frac{\sqrt{1-\lambda } }{\lambda }\biggl(\frac{2 \tan ^{-1}(\sqrt{1-\lambda}\, \text{sc}(\phi,\lambda ))}{\sqrt{1-\lambda}}\nonumber\\
& &-(2+\lambda) \text{am}(\phi, \lambda )\biggr).
\een
Due to the shift performed over $\phi$, the field solution of (\ref{phiprimeequalswphi}) becomes 
\begin{equation}
\phi_{\lambda}(y)=\text{sc}^{-1}\left(\frac{\tanh \left(\sqrt{1-\lambda^2} y\right)}{\sqrt{1+\lambda}},\lambda\right).
\end{equation}
Its shape can be observed in Fig.~\ref{fig3}, when we make the changes $x\to y$ and $\phi_{\lambda}(y)=\chi_{S,\lambda}(y)-K_{\lambda}$. Note that for $\lambda=1$ the field approaches a constant,  $\phi_{\lambda=1}(\pm\infty)=\pm\text{csch}^{-1}\left(\sqrt{2}\right)$.

With these ingredients we analytically solve the equation (\ref{afirstorder}), together with the boundary conditions $A(0)=A'(0)=0$, to find
\begin{widetext}
\begin{eqnarray}\label{warpfunction}
A(y)&=&-\frac{2}{3 \lambda} \left[\sqrt{1-\lambda}(2+\lambda){\cal F}\left(\sqrt{1+\lambda},\sqrt{1-\lambda ^2},y\right)-2 {\cal F}\left(\sqrt{\frac{1+\lambda}{1-\lambda }},\sqrt{1-\lambda ^2},y\right)\right], 
\end{eqnarray}
where
\begin{equation}
{\cal F}(b,a,x)=-x \cot ^{-1}(b)+\frac{i}{4 a} \left[\text{Li}_2\left(\frac{b+i}{i-b} e^{2 a x}\right)-\text{Li}_2\left(\frac{i-b}{b+i} e^{2 a x}\right)-\text{Li}_2\left(\frac{b+i}{i-b}\right)+\text{Li}_2\left(\frac{i-b}{b+i}\right)\right].
\end{equation}
\end{widetext}
Here $\text{Li}_2(y)$ is the polylogarithmic function. The shape of $e^{2A(y)}$ is depicted in Fig.~\ref{fig10}. We can observe that in the vicinity of the center the warp factor behaves like $A(|y|\simeq 0)\simeq (1-\lambda)y^2+\mathcal{O}(y^4)$, which implies that as  $\lambda\to 1$, the brane becomes less and less localized, becoming effectively delocalized for $\lambda=1$.  Meanwhile, in the asymptotic regime $|y|\to\infty$ the thick brane approaches AdS vacua with cosmological constant $\Lambda_5\sim -W(\phi(\pm\infty))^2$, and we have $A(|y|\to\infty)\simeq -\frac{1}{3}E_{S}(\lambda)|y|$, where we used the relations (\ref{Ebpsdeformed}) and (\ref{ES}). As $E_S(\lambda)$ decreases as $\lambda$ grows (starting at $\lambda=0$, where we have the sine-Gordon brane) the thick branes generated by this model approaches AdS vacua with smaller cosmological constant as $\lambda\to 1$. In particular, for $\lambda=1$ the brane seems to fill the entire space, so we have no graviton localization anymore at this point, with all the modes dispersed through the extra dimension. The braneworld scenario is destroyed in the limit $\lambda\to1$.

From equations (\ref{openee}) we find that the potential has to obey \eqref{potb} and one notes that it keeps the correct form for vacuum stability in gravitational theories \cite{boucher,townsend}. In particular, for $\lambda=1$ the potential $U(\phi)$ vanish, and the spacetime solution reaches a flat 5-dimensional geometry. At this point, only the dynamical term $\sim \partial^{a}\phi\partial_{a}\phi$ survives in the action (\ref{action}), having as solution the constants $\phi_{\lambda=1}(\pm\infty)$. The presence of two values possible for the scalar field in the flat background is due the original $\mathbb{Z}_2$-symmetry, which act as a memory related to the kind of thick brane system from which it is derived.

The energy density of the model is given by
\be\label{edb}
\rho(y)= e^{2 A}\left(W_{\phi}^2-\frac43 W^2\right),
\ee 
and its shape is depicted in Fig.~\ref{fig11}.  It is known that models derived from the first order equations \eqref{afirstorder} and \eqref{phiprimeequalswphi} have zero energy. It happens because we can rewrite the energy density as a total derivative, $\rho(y)=\frac{d}{dy}\left(W\,e^{2A}\right)$.  As  $W(\phi(\pm\infty))$ is finite and asymptotically the warp function falls off as $e^{-constant^2|y|}$,  the integral of $\rho(y)$ over all space must vanish. Here we see that when
$\lambda=0$, the energy density is well-concentrated around the origin and, as $\lambda$ increases, it becomes more and more diffuse until finally disappearing at $\lambda=1$.

The above results describe an interesting scenario, in which the parameter $\lambda$ may be used to control the physical properties of the 5-dimensional spacetime. If $\lambda$ increases from $0$ to unity, it may change the spacetime from a braneworld model with a single extra spatial dimension of infinite extent to a 5-dimensional Minkowski spacetime with no graviton localization.

\subsection{Metric fluctuations}

\begin{figure}[t]
\centerline{\includegraphics[height=14em]{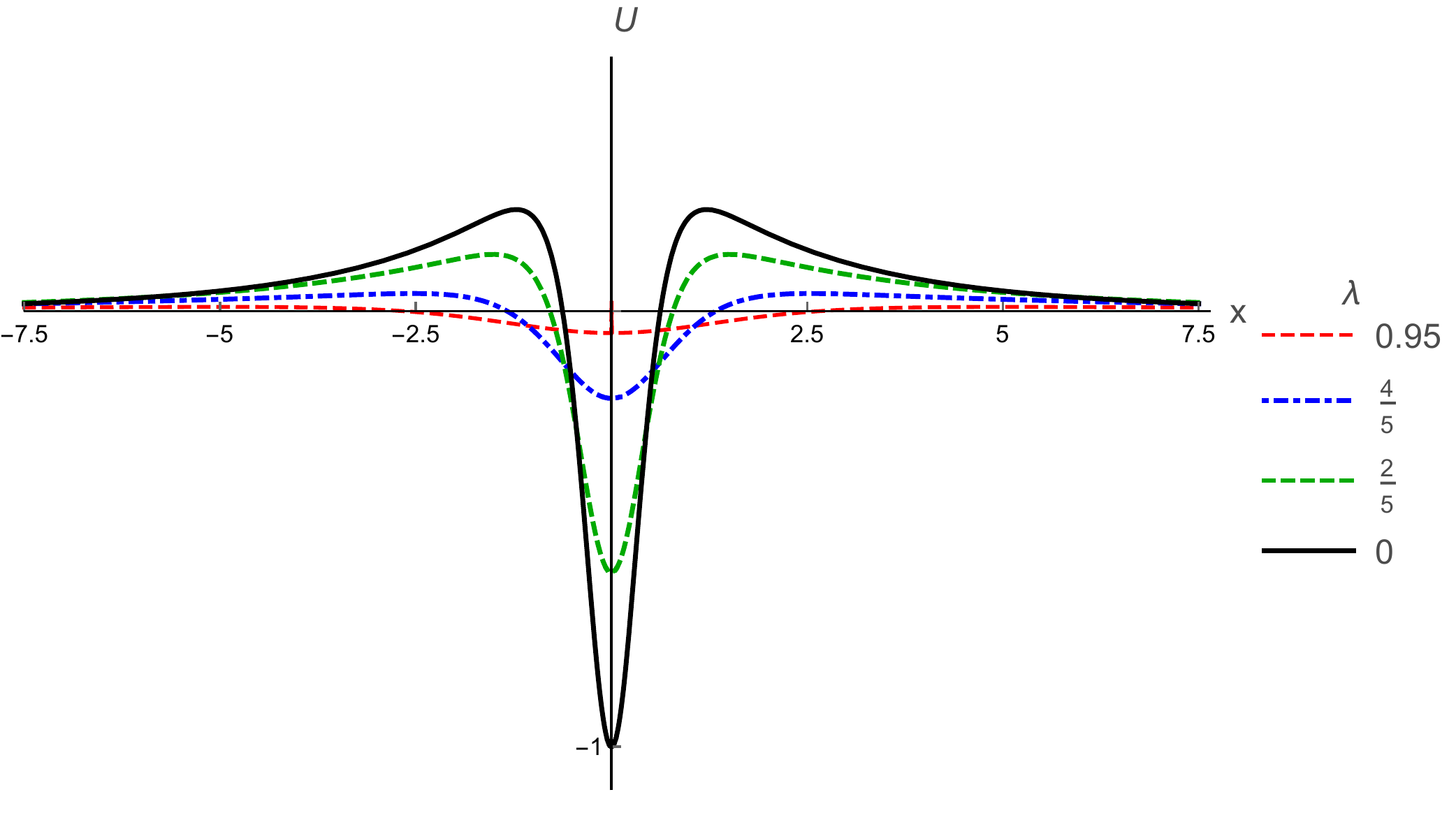}}
\caption{The stability potential (\ref{sp}) for some values of $\lambda$. Here we observe the stability potential of the gravitational sector  considerably decreases  when $\lambda\to1$, although it always maintains its volcano shape, implying stability for  gravity localization. For $\lambda=1$, equation (\ref{sp}) vanish.
 }\label{fig12}
\end{figure}

In this section we analyze the stability of the gravitational sector. For this, we perform a redefinition of variable $ dy^ 2 \rightarrow  e^{2A(z)} dz^2 $ in (\ref{metric}), which allows us to rewrite the metric in a conformally flat scenario $\tilde{g}_{ab}=e^{2A(z)}\eta_{ab}$. With a linear perturbation the metric becomes
\begin{equation}
ds^2=e^{2A(z)}\left(\eta_{ab}+h_{ab}\right)dx^a dx^b.
\end{equation}
In the  transverse-traceless  gauge ($\partial_{\mu}h^{\mu\nu}=0$ and $h_{\mu}^{\mu}=0$) the conformal Einstein tensor is ${\bar G}_{ab}=-\frac12\partial_c\partial^c h_{ab}$ and the linearized Einstein tensor is given by
\begin{eqnarray}
\nonumber G_{ab}^{(1)}&=&-\frac{1}{2}\partial_c\partial^c h_{ab}+3\Bigl[\partial_a A\partial_b A-\partial_a \partial_b A+\\
&~&+\frac{1}{2}A'h'_{ab}+\bar{g}_{ab}\left(\partial_c \partial^c A+\partial_c A\partial^c A\right)\Bigr].
\end{eqnarray}
In this way the $\mu\nu-$components of $G_{ab}^{(1)}$ are
\begin{equation}
G_{\mu\nu}^{(1)}=-\frac{1}{2}\partial_c\partial^c h_{\mu\nu}+\frac{3}{2}A'h'_{\mu\nu}-3\bar{g}_{\mu\nu}\left( A''+ A'^2\right).
\end{equation}
and the linearized energy-momentum tensor becomes
\begin{equation}
T_{\mu\nu}^{(1)}=-\frac{3}{2}\bar{g}_{\mu\nu}\left( A''+ A'^2\right)
\end{equation}
where the prime denotes the derivative in relation to variable $z$. When using linearized Einstein equations, $G_{\mu\nu}^{(1)}=2T_{\mu\nu}^{(1)}$, we obtain the equation for $h_{\mu\nu}$, which is $-\partial_c\partial^c h_{\mu\nu}+3A'h'_{\mu\nu}=0$. At last, the redefinition $H_{\mu\nu}=e^{-ipx}e^{3A/2}h_{\mu\nu}$ allows us to rewrite the equation for $h_{\mu\nu}$ as
\begin{equation}\label{eqH}
\left(\partial_z+\frac{3}{2}A'\right)\left(-\partial_z+\frac{3}{2}A'\right)H_{\mu\nu}=p^2 H_{\mu\nu}.
\end{equation}
Note that equation (\ref{eqH}) has the form of  a  supersymmetric Schr\"odinger equation, where the stability potential is 
\be\label{sp}
U(z)=\frac32 A^{\prime\prime}+ \frac94 A^{\prime 2}.
\ee
wich is  depicted in (\ref{fig12}). We can observe that (\ref{sp}) keep its volcano shape, as usual, but as $\lambda$ grows it becomes less expressive, and we have no stability potential for $\lambda=1$.  Equation (\ref{eqH}) has the factorized form $S^{+}S^{-}\psi=p^2\psi$, where  $S^{\pm}=\left(\pm\partial_z+3A'/2\right)$. In particular, the zero-energy solution of (\ref{eqH}) is given by $\psi_0(y)=e^{3A(y)/2}$, where the asymptotic behavior of $A(y)$ ensures gravity localization around the brane for $0\leq\lambda<1$. Since the Hermitian operator $S^{+}S^{-}$ is non-negative, we have no normalizable negative gravitons modes, and it ensures system stability.  At $\lambda = 1$, we find a flat space, so the stability analysis as presented here fails. In this case, what guarantees the stability of this space are the positive mass theorems on asymptotically flat spacetimes, which are valid for dimensions $\leq 7$ \cite{schoen}.
 
\subsection{ RG Flow}

As pointed out in \cite{skenderis,skenderis1}, by rewriting the metric (\ref{metric}) in the form
\begin{equation}
ds^2=u^2\eta_{\mu\nu}dx^\mu dx^\nu-\frac{1}{A'(y)^2}\frac{du^2}{u^2},
\end{equation}
where $u=e^{A(y)}$, allows us to interpret the function $u(y)$ as the renormalization scale of some Quantum Field Theory in the sense of Gauge/Gravity Duality \cite{maldacena}. Note that, due to the boundary conditions on the deformation factor, $u$ is  limited to the range $ [0,1] = [e^{A(\pm\infty)}, e^{A (0)}]$. Thus, domain wall solutions such as those presented here naturally lead us to confining regimes with a UV-cutoff located at $u = 1$ in the dual Field Theory. In this context, the solution $\phi(y(u))$ is identified with the running coupling of the system (see \cite{bbl,kiritsis} and references therein), so that the $\beta$-function is defined by
\begin{equation}
\beta(\phi)=u\frac{d\phi}{du}=-\frac{3}{2}\frac{W_\phi}{W}
\end{equation}
to describe the RG flow in the dual theory.

Denoting the critical points of (\ref{potb}) by $\phi_{\lambda,\infty}$, we can identify them with the zeros of $W_{\phi}$, which lead us to AdS (or flat) vaccua. Expanding the $\beta$-function around its critical point we have $\beta(\phi)\simeq \beta(\phi_{\lambda,\infty})+\beta'(\phi_{\lambda,\infty})(\phi-\phi_{\lambda,\infty})+\mathcal{O}\left((\phi-\phi_{\lambda,\infty})^2\right)$. Since in the critical points we have $\beta(\phi=\phi_{\lambda,\infty})=0$, we can find the following expression for the running coupling
\begin{equation}
\phi=\phi_{\lambda,\infty}+c u^{\beta'(\phi_{\lambda,\infty})}. 
\end{equation}
Note that if $\beta'(\phi_{\lambda,\infty})<0$,  $\phi_{\lambda,\infty}$ is a UV fixed point when $u\to\infty$, and if $\beta'(\phi_{\lambda,\infty})>0$, $\phi_{\lambda,\infty}$ is a IR fixed point for $u\to 0$. For $\beta'(\phi_{\lambda,\infty})=0$ we have a conformal theory.

For the model presented here, we have
\begin{equation}\label{dbeta}
\beta'(\phi_{\lambda,\infty})=\frac{3\lambda \text{cn}\left(\phi_{\lambda,\infty},\lambda \right) \text{sn}\left(\phi_{\lambda,\infty}, \lambda \right)}{\text{am}\left(\phi_{\lambda,\infty}, \lambda \right)-\frac{2}{\sqrt{1-\lambda}(2+\lambda)}\tan^{-1}\left(\sqrt{\frac{1-\lambda }{1+\lambda}}\right)}
\end{equation}
where $\phi_{\lambda,\infty}=\text{sc}^{-1}\left(\frac{1}{\sqrt{1+\lambda}},\lambda \right)$. Note that (\ref{dbeta}) is a monotonically increasing function of $\lambda$ and we have no divergences in the running coupling since $u\in[0,1]$, but  for any $\lambda$ the IR regime $(u\to 0)$ at $\phi=\phi_{\lambda,\infty}$ is well defined in the dual field theory, even for the 5d Minkowski setup. As a consequence, none of the solutions has a conformal dual field theory.

\section{Comments}
\label{end}

In this work we studied a sine-Gordon-like model, which is controlled by a real parameter that continuously connects the sine-Gordon and the vacuumless models. The model appears as a deformation of the $\phi^4$ model, and the real parameter is $\lambda$: for $\lambda=0$ one gets the standard sine-Gordon model and for $\lambda=1$ it reproduces the so-called vacuumless model. However, for $\lambda$ in the interval $(0,1)$ one gets a double sine-Gordon model, which contains two distinct topological sector, the large and the small sectors, which give rise to the large and small kinks, respectively.

As it was shown, in the 2-dimensional spacetime, the energy of the large kink in the large sector varies from $E_L(\lambda=0)=1$ to $E_L(\lambda=1)=2\pi$, and in the case of the small sector one gets $E_S(\lambda=0)=1$ and $E_S(\lambda=1)=0$. We then see that the small sector, which is degenerate to the large sector at $\lambda=0$, disappears as $\lambda=1$, with the large sector becoming the topological sector of the vacuumless model.  

In the 5-dimensional case, we considered a warped geometry with a single extra dimension of infinite extent and studied the new braneworld scenario described in the small sector. In this scenario, the brane energy density is such that the brane energy vanishes, independently of the value of
$\lambda$. If we see the model with $\lambda$ increasing from zero to unity, it then nicely describes a way to change a 5-dimensional warped geometry which is asymptotically AdS into a 5-dimensional Minkowski geometry. However, if $\lambda$ is supposed to run in the reverse sense, decreasing from unity to zero, the model could do the reverse, changing the 5-dimensional Minkowski geometry into a braneworld scenario with a warped geometry which is asymptotically AdS$_5$.

As we have shown, the model is stable under tensorial fluctuations in the metric and of current interest, so one should now investigate how it modifies Newton's law, and how fermion and gauge fields can be entrapped into the brane as $\lambda$ varies in the interval $[0,1]$. Another issue of current interest concerns the variation of $\lambda$: the present investigation cannot tell the value of $\lambda$, so one should search for this
considering other arguments. An interesting possibility could be to investigate the conformational entropy associated with the current braneworld model, to see how it behaves as $\lambda$ varies in the interval $[0,1]$. This has been recently investigated in other contexts in \cite{e1,e2,e3,e4} and in references therein, and may provide important guide towards the physical realization of gravity localization in the present model, since the entropy could perhaps suggest the better way $\lambda$ should vary, increasing or decreasing in the interval $[0,1]$.

\acknowledgements{We thank the Brazilian agencies CAPES and CNPq for financial support.}


\end{document}